\documentstyle[12pt,epsfig,graphics]{article}
\newcommand{\dfrac}{\displaystyle\frac}

\def\lint{\int\limits}

\def\dlint{\displaystyle\lint}
\def\dint{\displaystyle\int}
\def\mb#1{\mbox{\boldmath{$#1$}}}
\newlength{\oldparskip}
\begin{document}
\begin{titlepage}
\hfill\begin{tabular}{l}HEPHY-PUB 690/98\\UWThPh-1998-29\\June 1998
\end{tabular}\\[2.5cm]
\begin{center}
{\Large\bf BOUNDS ON SCHR\"ODINGER ENERGIES}\\[2ex]
{\large\bf An Illustration of the Application of Mathematica 3.0}\\[2.5cm]
{\Large\bf HAN Liang}\\[.2cm]Department of Modern Physics,\\University of Science and Technology 
of China,\\Hefei, China\\[0.7cm]
{\Large\bf Wolfgang {LUCHA}}\\[.2cm]Institut f\"ur Hochenergiephysik,\\\"Osterreichische Akademie 
der Wissenschaften,\\Nikolsdorfergasse 18, A-1050 Wien, Austria\\[0.7cm]
{\Large\bf MA Wen-Gan}\\[.2cm]Department of Modern Physics,\\University of Science and Technology 
of China,\\Hefei, China\\[0.7cm]
{\Large\bf Franz F.~{SCH\"OBERL}}\\[.2cm]Institut f\"ur Theoretische Physik,\\Universit\"at 
Wien,\\Boltzmanngasse 5, A-1090 Wien, Austria\\[1.7cm]
\normalsize\it
Based on invited basic lectures on numerical mathematics\\(presented by Franz F.~Sch\"oberl) at 
the Department of Modern Physics,\\University of Science and Technology of China, Hefei, China, 
March 1997.\\Supported by the Chinese Academy of Science and\\the Austrian Academic Exchange 
Service.
\end{center}
\end{titlepage}

\newpage

\section{Introduction}

Unfortunately, most physical problems cannot be solved analytically, which means that we are not 
able to give their solutions in closed mathematical form. Thus, we are forced to adopt numerical 
methods to obtain results. But, how to keep these numerical solutions under control? In fact, one 
has to know---in some sense---what will emerge numerically, what will be the approximate result. 
Consequently, the first step, before starting the computer, will be to get a feeling for the 
output. First, one will analyze the dimension of the output; secondly, one will try to ``guess'' 
the order of magnitude of the desired output; and, thirdly, one will improve this first 
``guess.''

In these lectures we illustrate, on various examples, how to derive, even for rather complicated 
problems, quantities like upper bounds on energy eigenvalues analytically, and we improve our 
results by numerical methods. In both cases, we shall demonstrate how to use Mathematica, 
described as ``a system for doing mathematics by computer'' \cite{mathemat97}.

First, we will discuss how to handle dimensions and how to apply scaling arguments to the 
Schr\"odinger equation. After introducing the variational procedure, we calculate upper bounds on 
the energy eigenvalues of the Schr\"odinger Hamiltonian, using ``paper and pencil'' as well as 
Mathematica 3.0 \cite{mathemat97}.

\begin{figure}[htb]
\begin{center}
\psfig{figure=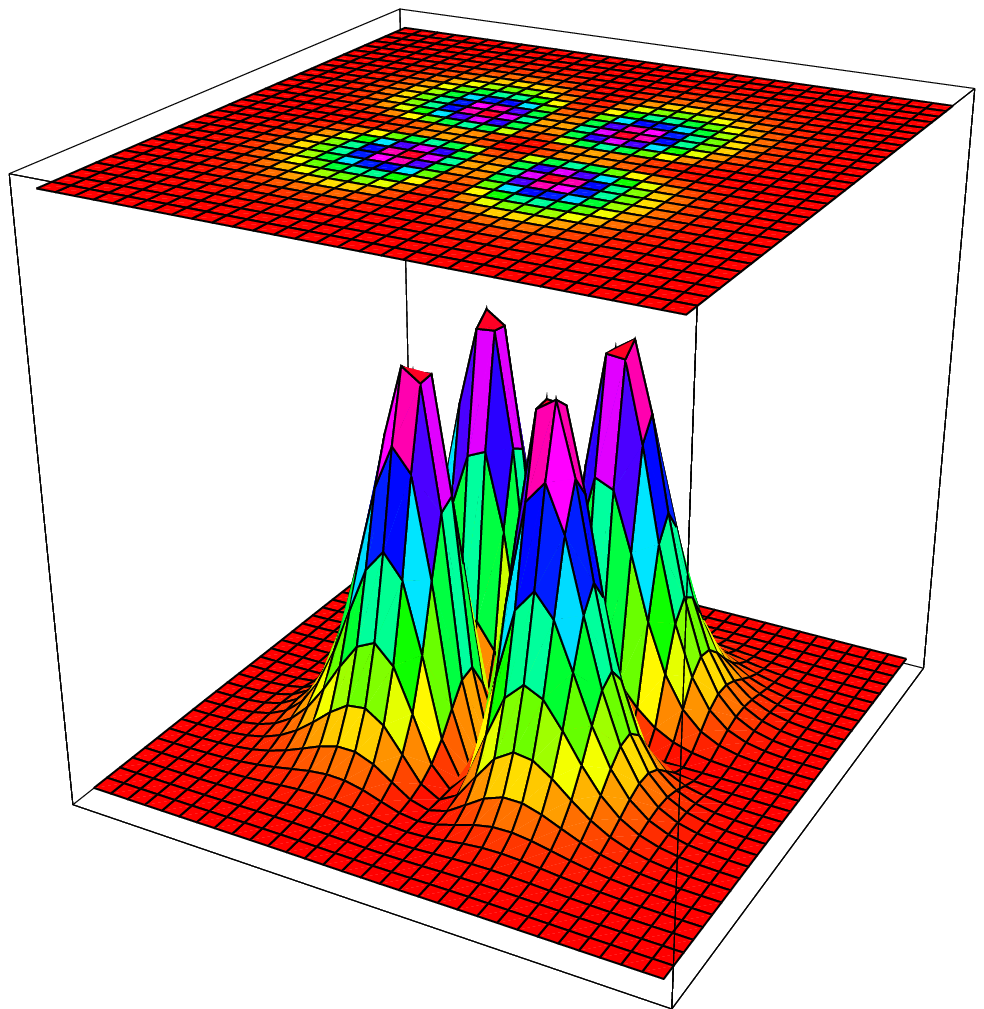,scale=0.8}
\caption{The eigenfunction of the H atom, for radial quantum number $n_{\rm r}=0$, relative 
orbital angular momentum $\ell=2$, and its projection $m=1$}\label{fig:Hydrogen021}
\end{center}
\end{figure}

One of the main advantages of the use of Mathematica is the capability to represent results very 
easily in graphical form. This feature is illustrated in Fig.~\ref{fig:Hydrogen021} for a 
particular D-wave eigenfunction of the hydrogen atom.

\newpage

\section{Dimensional Reasoning}

We all know that it is a horrible task to handle dimensions. In particle physics, at~least, it is 
customary to use a system of ``natural'' units defined by 
$$
\fbox{$\hbar=c=1$}\ .
$$
Using these units, we are able to express every dimensional physical quantity in, e.g., units of 
energy $E$. Clearly,
$$
\mbox{Dim}[\hbar]=\mbox{Dim}[c]=1=\mbox{Dim}[E^0]\ .
$$
Then
$$
\mbox{Dim}[E\,t]=\mbox{Dim}[E]\,\mbox{Dim}[t]=\mbox{Dim}[\hbar]=1
$$
and, consequently, the dimension of time $t$ is the inverse of the dimension of energy:
$$
\fbox{$\mbox{Dim}[t]=\displaystyle{1\over\mbox{Dim}[E]}=\mbox{Dim}[E^{-1}]$}\ .
$$
From the well-known relation between the mass $m$ and the rest energy $E_0$ of a particle,
$$
E_0=m\,c^{2}\ ,
$$
the dimension of any mass $m$ is
$$
\fbox{$\mbox{Dim}[m]=\mbox{Dim}[E]$}\ .
$$
Because of
$$
\mbox{Dim}[p]=\mbox{Dim}[m\,c]=\mbox{Dim}[m]\ ,
$$
the dimension of any momentum $p$ is also
$$
\fbox{$\mbox{Dim}[p]=\mbox{Dim}[E]$}\ .
$$
The dimension of any spatial coordinate $x$ is obtained from
$$
\mbox{Dim}[p\,x]=\mbox{Dim}[\hbar]=1
$$
as
$$
\fbox{$\mbox{Dim}[x]={\displaystyle{1\over\mbox{Dim}[p]}}=\mbox{Dim}\left[E^{-1}\right]$}\ .
$$
We are now able to express any physical quantity in units of energy. For instance, let us 
calculate the dimension of the Schr\"odinger wave function $\Psi$. The Schr\"odinger wave 
function $\Psi({\bf x})$ in configuration-space representation is usually normalized according~to
$$
\int {\rm d}^3x\,\Psi^*({\bf x})\,\Psi({\bf x})=1\ ,
$$
which entails for the dimensions
$$
\mbox{Dim}[x^3\,\Psi^2]=1\ ,
$$
hence,
$$
\mbox{Dim}[\Psi^2]=\frac{1}{\mbox{Dim}[x^3]}\ ,
$$
implying
$$
\fbox{$\mbox{Dim}[\Psi]=\mbox{Dim}[E^{3/2}]$}\ .
$$
Let us apply, just as an example, this method of counting dimensions to the evaluation of the 
Fourier transform of $1/{\bf p}^2$. As we know, the detailed calculation needs some~care, and, 
involving a distribution, is not completely trivial. The quantity to be evaluated~is
$$
\int{\rm d}^3p\,\frac{e^{-i\bf{p}{\bf\cdot}\bf{x}}}{{\bf p}^2}\ .
$$
Obviously, the only free parameter in the above integrand is $\bf{x}$, and since the integrand 
contains the rotationally invariant scalar product $\bf{p}{\bf \cdot }\bf{x}$, the integral must 
be a function of merely the modulus $\left|\bf{x}\right|$ of the (spatial) coordinates ${\bf x}$. 
Now, let us count dimensions:
$$
\mbox{Dim}\left[\frac{p^3}{p^2}\right]=\mbox{Dim}[p]=\mbox{Dim}[E]=\frac{1}{\mbox{Dim}[x]}\ .
$$
Therefore, the integral has to be proportional to the inverse of $|{\bf x}|$:
$$
\int{\rm d}^3p\,\frac{e^{-i\bf{p}{\bf\cdot}\bf{x}}}{{\bf p}^2}=\frac{A}{\left|\bf{x}\right|}\ .
$$
The constant of proportionality $A$ is now easily determined by applying the Laplacian 
$\Delta\equiv\mb\nabla\mb\cdot\mb\nabla$ to both sides of this equation, and by using the rather 
well-known relations
$$
\Delta\frac{1}{\left|\bf{x}\right|}=-4\pi\,\delta^{(3)}(\bf{x})
$$
as well as
$$
\int{\rm d}^3p\,e^{-i\bf{p}{\bf\cdot}\bf{x}}=(2\pi)^3\,\delta^{(3)}(\bf{x})\ .
$$
This procedure results in
$$
\Delta\int{\rm d}^3p\,\frac{e^{-i{\bf p}{\bf\cdot}{\bf x}}}{{\bf p}^2}=
-\int{\rm d}^3p\,e^{-i{\bf p}{\bf\cdot}{\bf x}}=-(2\pi)^3\,\delta^{(3)}({\bf x})=
\Delta\frac{A}{\left|{\bf x}\right|}=-4\pi\,A\,\delta^{(3)}({\bf x})\ .
$$
From this relation we read off for the constant of proportionality $A$ in the above~ansatz:
$$
A=\frac{(2\pi)^3}{4\pi}=2\pi^2\ .
$$
Just by counting dimensions and by using general physical arguments, we find, without any 
difficult integrations in the complex plane,
$$
\fbox{$\dint{\rm d}^3p\,{\displaystyle{e^{-i\bf{p}{\bf\cdot}\bf{x}}\over{\bf p}^2}}
={\displaystyle{2\pi^2\over\left|\bf{x}\right|}}$}\ .
$$
Along the lines sketched above, we will show how to solve rather complicated problems, just by 
applying general theorems and rules.

\section{The Schr\"odinger Equation}

Solving the Schr\"odinger equation is, in general, a somewhat tedious task. However, as long as 
we are merely interested in the dependence of the energy eigenvalues $E$ on the parameters 
entering in the Schr\"odinger equation, we do not have to solve the equation at all. We can adopt 
physical arguments only, arguments like the scaling behaviour of the Schr\"odinger equation.

\subsection{Scaling Behaviour}

The time-independent Schr\"odinger equation for a generic power-law central potential, i.e., a 
potential of the form $V(r)=a\,r^n$, which depends merely on the radial coordinate $r\equiv|{\bf 
x}|$, reads in configuration space, abbreviating the Laplacian again by 
$\Delta\equiv\mb\nabla\mb\cdot\mb\nabla$,
\begin{equation}
\fbox{$\left(-\dfrac{\Delta}{2\,\mu}+a\,r^{n}\right)\Psi({\bf x})=E\,\Psi({\bf x})$}\ ,
\label{eq:schrodi1}
\end{equation}
with the reduced mass $\mu$ of the two-particle bound system under consideration defined by
$$
\mu\equiv\frac{m_1\,m_2}{m_1+m_2}\ .
$$
Let us start by scaling, just for fun, the coordinates ${\bf x}$ by some (arbitrarily 
chosen)~scale parameter~$\kappa$,
\begin{equation}
{\bf x}=\kappa\,{\mb\rho}\ ,
\label{eq:scale1}
\end{equation}
which gives, because of
$$
\Delta=\frac{\Delta_\rho}{\kappa^2}
$$
and
$$
r^n=\kappa^n\,\rho^n\ ,
$$
the scaled (time-independent) Schr\"odinger equation
$$
\left(-\frac{\Delta_\rho}{2\,\mu\,\kappa^2}+a\,\kappa^n\,\rho^n\right)\Psi(\kappa\,{\mb\rho})
=E\,\Psi(\kappa\,{\mb\rho})\ .
$$
Multiplying now the above equation by $2\,\mu\,\kappa^2$ leads to
$$
\left(-\Delta_\rho+2\,\mu\,a\,\kappa^{2+n}\,\rho^n\right)\Psi=2\,\mu\,\kappa^2\,E\,\Psi\ .
$$
Apart from our pure amusement, what might be the deeper reason for this re-scaling? Obviously, we 
can choose $\kappa$ in such a way that the factor $2\,\mu\,a$ disappears:
$$
\kappa^{2+n}=\frac{1}{2\,\mu\,a}\ ,
$$
and thus
\begin{equation}
\kappa=\left(\frac{1}{2\,\mu\,a}\right)^{1/(2+n)}\ .
\label{eq:kappa}
\end{equation}
For this particular choice of $\kappa$, the scaled Schr\"odinger equation is cast into the form
\begin{equation}
\fbox{$\left(-\Delta_\rho+\rho^n\right)\Psi=\varepsilon\,\Psi$}\ ,
\label{eq:scaledschrodi}
\end{equation}
with $\varepsilon$ some (dimensionless) number which has to be determined by solving the scaled 
Schr\"odinger equation. Note that this scaled Schr\"odinger equation is dimensionless. We have 
thus managed to separate physics from ``pure'' mathematics by the identification
$$
2\,\mu\,\kappa^2\,E\equiv\varepsilon\ .
$$
Inserting Eq. (\ref{eq:kappa}) shows us how the energy eigenvalue $E$ depends on the 
parameters~$a$, $\mu$, and $n$ entering in the Schr\"odinger equation:
\begin{equation}
\fbox{$E=\left[{\displaystyle{a^2\over\left(2\,\mu\right)^n}}\right]^{1/(2+n)}\,\varepsilon$}\ .
\label{eq:scaledenergy}
\end{equation}
Now we are in the position to discuss the physical behaviour of the energy eigenvalues $E$ for 
various power-law potentials without solving explicitly the Schr\"odinger equation.

\begin{description}
\item{\bf Coulomb potential} ($n=-1$):
$$
E=2\,\mu\,a^2\,\varepsilon\ . 
$$
This energy eigenvalue is proportional to the mass and to the coupling constant squared. The 
energy eigenvalue $E$ has to be proportional to the mass because the only parameter with 
dimension of energy entering in the Schr\"odinger equation is the mass $\mu$. However, we cannot 
conclude from dimensional considerations~only how the energy eigenvalue $E$ depends on the (fine-
structure) coupling constant~$a$.
\item{\bf Linear potential} ($n=1$):
\begin{equation}
E=\left(\frac{a^2}{2\,\mu}\right)^{1/3}\,\varepsilon\ .
\label{eq:scaledlinear}
\end{equation}
In contrast to the Coulomb potential, this energy eigenvalue $E$ is proportional~to the inverse 
of the mass $\mu$ to the power $1/3$.
\item{\bf Logarithmic potential} ($n=0$): Because of the (easily to check) identity
$$
\ln r=\lim_{n\rightarrow 0}\frac{r^n-1}{n}\ ,
$$
the logarithmic-potential energy eigenvalue $E$ is obtained for $n=0$ in Eq. 
(\ref{eq:scaledenergy}):
$$
E=a\,\varepsilon\ .
$$
Note that in this case the energy eigenvalue $E$ is independent of the mass. That means, that, 
for instance, the difference of energy eigenvalues of excited states~is the same even for 
different masses $\mu$.
\end{description}
We have learned how to find the parameter dependence of the energy eigenvalues $E$~of the 
Schr\"odinger equation without solving this differential equation explicitly. Thus we are able to 
check numerical solutions by calculating ratios which are independent of~$\varepsilon$, like for, 
e.g., the linear potential,
$$
\frac{E_1}{E_2}=\left[\left(\frac{a_1}{a_2}\right)^2\,\frac{\mu_2}{\mu_1}\right]^{1/3}\ .
$$
In this way, we are able to get a feeling for the accuracy of any numerical calculation. 

\subsection{The Variational Method: A Derivation of Upper Bounds}

Since we are interested merely in an explicit application of the variational method, we omit all 
mathematically subtle discussions and explanations here. A discussion of the history of 
inequalities and variational methods for eigenvalue problems can be found~in Ref. 
\cite{weinstein72}, and some applications are given in Ref. \cite{flamm1982}. In Ref. \cite{reed-
simon}, a rigorous discussion of the problem can be found. For our purpose, in the frame of these 
lecture notes, it~is sufficient to know that, for any given Hamiltonian $H$ with 
eigenvalues~$E_k$, $k=1,2,\dots$, upper bounds $\widehat{E}_k$ on $E_k$ may be obtained by 
computing the matrix elements of $H$ with respect to some suitably chosen set of orthogonalized 
trial functions $\Psi_k$ which depend on some variational parameter $\lambda$. These upper bounds 
can be improved by minimizing the resulting upper bounds $\widehat{E}_k\left(\lambda\right)$ on 
the ``true'' energy eigenvalues with respect to the variational parameter $\lambda$. In order to 
obtain also upper bounds for the radially excited states, one has to diagonalize the 
corresponding energy matrix $E_{ij}$. The detailed steps of this general procedure are:
\begin{enumerate}
\item Choose a set of mutually orthogonal trial states $|\Psi_i(\lambda)\rangle$, i.e., 
$\left\langle\Psi_i(\lambda)|\Psi_j(\lambda)\right\rangle=\delta_{ij}$.
\item Determine the matrix elements of the Hamiltonian $H$ using these trial functions:
$$
E_{ij}(\lambda)\equiv\left\langle\Psi_i(\lambda)\left|H\right|\Psi_j(\lambda)\right\rangle\ .
$$
\item Determine the roots of the characteristic equation
\begin{equation}
\det\left[E_{ij}(\lambda)-\widehat{E}(\lambda)\,\delta_{ij}\right]=0\ .
\label{eq:determinant}
\end{equation}
\item These roots $\widehat{E}(\lambda)$ are upper bounds for any variational parameter $\lambda$ 
you choose.
\item You may want to improve the upper bounds by minimizing $\widehat{E}(\lambda)$ with 
respect~to $\lambda$:
\begin{equation}
\frac{\partial\widehat{E}(\lambda)}{\partial\lambda}=0\quad\leadsto\quad\lambda_{\rm{min}}\ . 
\label{eq:lambdamin1}
\end{equation}
\item In this way, you obtain the minimal upper bound $\widehat{E}_k(\lambda_{\rm{min}})$ on 
$E_k$ (in your~chosen sector of Hilbert space):
\begin{equation}
E_k\leq\widehat{E}_k(\lambda_{\rm{min}})\ .
\label{eq:lambdamin2}
\end{equation}
\end{enumerate}
We are now in a position to apply the above concepts to various power-law potentials.

\subsubsection{Ground States}

Now, let us start by considering ground states. The prototype for all studies of this~kind is the 
Coulomb problem. The Hamiltonian $H$ under consideration here is simply given by
\begin{equation}
H=\frac{{\bf p}^2}{2\,\mu}-\frac{\alpha}{r}\ ,
\label{eq:schrodiham}
\end{equation}
with the electromagnetic fine structure constant $\alpha$. First, we have to choose some trial 
function. Let us take, for ``didactic'' reasons, the hydrogen ground-state eigenfunction (of 
course, you can also choose, e.g., Gaussian trial functions)
$$
\Psi(r,\lambda)=N\,e^{-\lambda\,r}\ ,\quad\lambda^\ast=\lambda>0\ ,
$$
with the variational parameter, $\lambda$, and the normalization factor, $N$, to be determined. 
Secondly, calculate the normalization factor $N$:
$$
\dint{\rm d}^3x\,\Psi^*\,\Psi=1=N^2\dint{\rm d}^3x\,e^{-2\,\lambda\,r}=
N^2\,4\pi\dlint_0^\infty{\rm d}r\,r^2\,e^{-
2\,\lambda\,r}=N^2\,4\pi\,\frac{\Gamma(3)}{(2\,\lambda)^3}
\ ,
$$
where we have used
\begin{equation}
\fbox{$\dlint_0^\infty{\rm d}r\,r^n\,e^{-
\lambda\,r}={\displaystyle{\Gamma(n+1)\over\lambda^{n+1}}}$}
\ .
\label{eq:gammafunction}
\end{equation}
In Mathematica \cite{mathemat97}, the procedure is the following:
\noindent\vspace*{3mm}\hrule
\begin{center}{\sl MATHEMATICA}\end{center}
\hrule\vspace*{3mm}
\noindent{\sl Integration:}

\bigskip\noindent{\tt In[1]:= Integrate[r\symbol{94}n Exp[-lambda r],\{r,0,Infinity\}]}

\bigskip\noindent{\tt Out[1]= If[Re[lambda] \mbox{$>$} 0 \&\& Re[n] \mbox{$>$} -1,
lambda}$^{\mbox{\tt (-1-n)}}${\tt\ Gamma[1+n], }${\tt \dint_0^{\infty}
{\displaystyle {\mbox{r}^n \over \mbox{e\symbol{94}(lambda r)}}}} ${\tt \ dr]}

\vspace*{3mm}\hrule

\bigskip\noindent
This output has to be interpreted in the following way: If ${\Re\,\lambda>0}$ and ${\Re\,n>-1}$, 
then the result of this integration reads ${\lambda}^{-1-n}\,{\Gamma(1+n)}$, otherwise, the input 
is returned as
$$
\dint_0^\infty\dfrac{r^n}{e^{\lambda\,r}}\,{\rm d}r\ ,
$$
which means that Mathematica is not able to find a solution. The normalization factor is thus
$$
N=\frac{\lambda^{3/2}}{\sqrt{\pi}}\ ,
$$
and the normalized trial function reads
$$
\Psi(r,\lambda)=\frac{\lambda^{3/2}}{\sqrt{\pi}}\,e^{-\lambda\,r}\ .
$$
In order to be on the safe side, we check the dimensions of this expression on both~sides of this 
relation. As we have already discussed before, the left-hand side has dimension
$$
\mbox{Dim}[ E^{3/2}]\ .
$$
For the exponential $\exp(-\lambda\,r)$ to make sense at all, $\lambda\,r$ has to be 
dimensionless, which requires
$$
\mbox{Dim}[\lambda\,r]=1\ ,
$$
leading to
$$
\mbox{Dim}[\lambda]=\frac{1}{\mbox{Dim}[r]}=\mbox{Dim}[E]\ ,
$$
and, therefore,
$$
\mbox{Dim}[\Psi]=\mbox{Dim}[E^{3/2}]=\mbox{Dim}[\lambda^{3/2}]\ .
$$
Obviously, the dimensions, at least, are correct. Now we will show how the above steps are 
defined and calculated using Mathematica. We give the printout as it is returned.
\noindent\vspace*{3mm}\hrule
\begin{center}{\sl MATHEMATICA}\end{center}
\hrule\vspace*{3mm}
\noindent
{\sl Defining the trial function:}

\bigskip\noindent{\tt In[2]:= psi[r\_,lambda\_] := Exp[-lambda r]}

\bigskip\noindent
{\sl Integration:}

\bigskip\noindent{\tt In[3]:= 4 Pi Integrate[r\symbol{94}2 
psi[r,lambda]\symbol{94}2,\{r,0,Infinity\}]}

\bigskip\noindent{\tt Out[3]= 4}$\pi ${\tt \ If[Re[lambda] \mbox{$>$} 0, }$
{\tt \displaystyle {1\over\mbox{\tt 4 lambda}^{3}}}
${\tt , }$\dint_0^\infty${\tt e}$^{\mbox{\tt -2 lambda r}}${\tt \ r}$^{2}$ {\tt dr]}

\vspace*{3mm}\hrule

\bigskip\noindent
This output has to be read as: If ${\Re\,\lambda>0}$, then the result is
$$
\frac{4\pi}{4\,\lambda^{3}}\ ,
$$
otherwise, Mathematica returns the input,
$$
4\pi\int_0^\infty e^{-2\,\lambda\,r}\,r^2\,{\rm d}r\ ,
$$
indicating that Mathematica is not able to evaluate this input. The complete result~is 
$$
N^2\,\frac{4\pi}{4\,\lambda^3}=1\quad\leadsto\quad N=\sqrt{\dfrac{\lambda^3}{\pi}}\ .
$$
Next, we have to evaluate the expectation values of the kinetic energy with respect~to our trial 
function. Since we are only interested in the energy of the ground state,~which is always 
characterized by $\ell = 0$, for our purpose the Laplacian $\Delta$ is effectively given~by
$$
\Delta=\frac{{\rm d}^2}{{\rm d}r^2}+\frac{2}{r}\,\frac{\rm d}{{\rm d}r}\ .
$$
Its expectation value thus reads
\begin{eqnarray*}
\dint{\rm d}^3x\,\Psi^*(r,\lambda)\,\Delta\Psi(r,\lambda)
&=&\frac{\lambda^3}{\pi}\int{\rm d}^3x\,e^{-\lambda\,r}\left(\frac{{\rm d}^2}{{\rm 
d}r^2}+\frac{2}{r}\,\frac{\rm d}{{\rm d}r}\right)e^{-\lambda\,r}\\[1ex]
&=&\frac{\lambda^3}{\pi}\,4\pi\dlint_0^\infty{\rm d}r\left(r^2\,\lambda^2-2\,r\,\lambda\right)
e^{-2\,\lambda\,r}\\[1ex]
&=&4\,\lambda^3\left[\lambda^2\,\frac{\Gamma(3)}{(2\,\lambda)^3}-
2\,\lambda\,\frac{\Gamma(2)}{(2\,\lambda)^2}\right]\\[1ex]
&=&4\,\lambda^3\left(-\frac{1}{4\,\lambda}\right)\\[1ex]
&=&-\lambda^2\ .
\end{eqnarray*}
Again we will check the dimensions and we will omit from now on the symbol Dim$[\dots]$:
$$
x^3\,\Psi\,\frac{1}{x^2}\,\Psi\sim E^{-3}\,E^{3/2}\,E^2\,E^{3/2}=E^2\sim\lambda^2\ .
$$
Fortunately, all dimensions are O.K. The expectation value of the kinetic energy is~thus 
\begin{equation}
\fbox{$\left\langle{\displaystyle{{\bf p}^2\over 2\,\mu}}\right\rangle=\dint{\rm 
d}^3x\,\Psi^*(r,\lambda)\left(-\dfrac{\Delta}{2\,\mu}\right)\Psi(r,\lambda)=
{\displaystyle{\lambda^2\over 2\,\mu}}$}\ .
\label{eq:kineticenergy}
\end{equation}
Already knowing how to interpret Mathematica output, below we just give the output. 
\noindent\vspace*{3mm}\hrule
\begin{center}{\sl MATHEMATICA}\end{center}
\hrule\vspace*{3mm}
\noindent
{\sl Defining the trial function:}

\bigskip\noindent{\tt In[4]:= psi[r\_,lambda\_] := lambda\symbol{94}(3/2)/Sqrt[Pi] Exp[-lambda 
r]}

\bigskip\noindent{\sl Effective Laplacian for vanishing orbital angular momentum:}

\bigskip\noindent{\tt In[5]:= g[r\_,lambda\_] := D[psi[r,lambda],\{r,2\}]+2/r 
D[psi[r,lambda],\{r,1\}]}

\bigskip\noindent{\sl Integration:}

\bigskip\noindent{\tt In[6]:= 4 Pi Integrate[r\symbol{94}2 psi[r,lambda] 
g[r,lambda],\{r,0,Infinity\}]}

\bigskip\noindent{\tt Out[6]= 4}$\pi$ $\left({\tt -}
{\tt\displaystyle{\mbox{\tt lambda}^2\over 4\pi}}\right)$

\vspace*{3mm}\hrule

\bigskip\noindent
The expression {\tt D[psi[r,lambda],\{r,$n$\}]} is the $n$th derivative of ${\Psi(r,\lambda)}$ 
with respect to $r$. The expectation value of a power-law potential $V(r)=a\,r^n$ reads 
\begin{eqnarray*}
\dint{\rm d}^3x\,\Psi^*(r,\lambda)\,V(r)\,\Psi(r,\lambda)
&=&\frac{\lambda^3}{\pi}\,4\pi\,a\dlint_0^\infty{\rm d}r\,r^{n+2}\,e^{-2\,\lambda\,r}\\[1ex]
&=&\frac{\lambda^3}{\pi}\,4\pi\,a\,\frac{\Gamma(n+3)}{(2\,\lambda)^{n+3}}\ ,
\end{eqnarray*}
that is,
\begin{equation}
\fbox{$\left\langle V(r)\right\rangle\equiv\dint{\rm 
d}^3x\,\Psi^*(r,\lambda)\,V(r)\,\Psi(r,\lambda)=4\,a\,\lambda^3\,
{\displaystyle{\Gamma(n+3)\over(2\,\lambda)^{n+3}}}$}\ .
\label{eq:expectpotential}
\end{equation}
For checking again the dimensions, we need the dimension of the coupling constant $a$:
$$
V=a\,r^n\sim a\,E^{-n}\sim E
$$
implies
$$
a\sim E^{n+1}\ .
$$
Without surprise, we realize that the dimensions in Eq. (\ref{eq:expectpotential}) are O.K.:
$$
a\,\lambda^{-n}\sim E^{n+1}\,E^{-n}=E\ .
$$

\newpage
\noindent\vspace*{3mm}\hrule
\begin{center}{\sl MATHEMATICA}\end{center}
\hrule\vspace*{3mm}
\noindent{\sl Integration:}

\bigskip\noindent{\tt In[7]:= Integrate[r\symbol{94}(n+2) Exp[-2 lambda r],\{r,0,Infinity\}]} 

\bigskip\noindent{\tt Out[7]= 2}$^{\mbox{\tt -3-n}}$ {\tt lambda}$^{\mbox{\tt -3-n}}${\tt\ 
Gamma[3+n]}

\vspace*{3mm}\hrule

\bigskip\noindent
Combining Eqs. (\ref{eq:kineticenergy}) and (\ref{eq:expectpotential}) yields the variational 
energy
\begin{equation}
E(\lambda)=\frac{\lambda^2}{2\,\mu}+\frac{a}{2}\,\frac{\Gamma(n+3)}{(2\,\lambda)^n}\ .
\label{eq:totalenergy}
\end{equation}
This energy is always an upper bound to the ``true'' energy for any $\lambda>0$: 
$E_{\rm{true}}\leq E(\lambda)$. We are easily able to improve this upper bound $E(\lambda)$ by 
determining that value $\lambda_{\rm{min}}$~of the variational parameter $\lambda$ which 
minimizes $E(\lambda)$; clearly, $\lambda_{\rm{min}}$ is then obtained from
$$
\frac{\partial E(\lambda)}{\partial\lambda}=0\ .
$$
The derivative of Eq. (\ref{eq:totalenergy}) with respect to $\lambda$ entails the requirement
$$
\frac{\partial E(\lambda)}{\partial\lambda}=\frac{\lambda}{\mu}-
a\,n\,\frac{\Gamma(n+3)}{(2\,\lambda)^{n+1}}=0\ ,
$$
which is solved by
\begin{equation}
\lambda_{\rm{min}}=\left[\frac{a\,n\,\mu\,\Gamma(n+3)}{2^{n+1}}\right]^{1/(n+2)}\ , 
\label{eq:lambdamin}
\end{equation}
which value, when re-inserted into Eq. (\ref{eq:totalenergy}), entails, in turn, the improved 
upper bound
\begin{equation}
\fbox{$E_{\rm var}\equiv E\left(\lambda_{\rm{min}}\right)=
\dfrac{1}{2}\left({\displaystyle{1\over\mu}}\right)^{n/(n+2)}
\left[{\displaystyle{a\,n\,\Gamma(n+3)\over 
2^{n+1}}}\right]^{2/(n+2)}\left(1+{\displaystyle{2\over n}}\right)$}\ .
\label{eq:improvede}
\end{equation}
\noindent\vspace*{3mm}\hrule
\begin{center}{\sl MATHEMATICA}\end{center}
\hrule\vspace*{3mm}

\noindent{\sl Defining the function $E(\lambda)$:}

\bigskip\noindent{\tt In[8]:= e[lambda\_] := lambda\symbol{94}2/(2 mu) + a/2 Gamma[n+3]/(2
lambda)\symbol{94}n}

\bigskip\noindent{\sl Differentiation with respect to the parameter $\lambda$ (note that\/ {\tt 
D[e[lambda],\{lambda,1\}]} is equivalent to\/ {\tt D[e[lambda],lambda]}):}

\bigskip\noindent{\tt In[9]:= D[e[lambda],lambda]}

\bigskip\noindent{\tt Out[9]= }${\displaystyle{\mbox{\tt lambda}\over\mbox{\tt mu}}}
${\tt\ - 2}$^{\mbox{\tt -1-n}}${\tt\ a lambda}$^{\mbox{\tt -1-n}}${\tt\ n Gamma[3+n]}

\bigskip\noindent{\sl Solving the equation for $\lambda_{\rm{min}}$:}

\bigskip\noindent{\tt In[10]:= Solve[}${\displaystyle{\mbox{\tt lambda}\over\mbox{\tt mu}}}
${\tt\ - 2}$^{\mbox{\tt -1-n}}${\tt\ a lambda}$^{\mbox{\tt -1-n}}${\tt\ n Gamma[3+n] == 
0,lambda]}

\bigskip\noindent{\tt Out[10]= }$\left\{\mbox{\tt lambda}\rightarrow\left(\mbox{\tt 2}^{\mbox{\tt 
-1-n}}\mbox{\tt\ a mu n Gamma[3+n]}\right)^{\frac{\mbox{\tt 1}}{\mbox{\tt 2+n}}}\right\}$

\bigskip\noindent{\sl Calculation of $E\left(\lambda_{\rm{min}}\right)$:}

\bigskip\noindent{\tt In[11]:= e[(2}$^{\mbox{\tt -1-n}}${\tt\ a mu n 
Gamma[3+n])}$^{\frac{\mbox{\tt 1}}{\mbox{\tt 2+n}}}${\tt ]}

\bigskip\noindent{\tt Out[11]= }${\displaystyle{\mbox{\tt (2}^{\mbox{\tt -1-n}}\mbox{\tt\ a mu n 
Gamma[3+n])}^{\frac{\mbox{\tt 2}}{\mbox{\tt 2+n}}}\over\mbox{\tt 2 mu}}} ${\tt\ + 2}$^{\mbox{\tt 
-1-n}}${\tt\ a Gamma[3+n]}$\left(\left(\mbox{\tt 2}^{\mbox{\tt -1-n}}\mbox{\tt\ a mu n 
Gamma[3+n]}\right)^{\frac{\mbox{\tt 1}}{\mbox{\tt 2+n}}}\right)^{\mbox{\tt -n}}$

\bigskip\noindent{\sl Note: The command\/ {\tt PowerExpand[{\it expr}]} expands all powers of 
products and powers. {\tt \%} stands for the last expression (here\/ {\tt Out[6]}).}

\bigskip\noindent{\tt In[12]:= PowerExpand[\%]}

\bigskip\noindent{\tt Out[12]= 2}$^{\frac{\mbox{\tt -4-3n}}{\mbox{\tt 2+n}}}${\tt\ 
a}$^{\frac{\mbox{\tt 2}}{\mbox{\tt 2+n}}}${\tt\ mu}$^{\tt -\frac{\mbox{\tt n}}{\mbox{\tt 
2+n}}}${\tt\ n}$^{\tt -\frac{\mbox{\tt n}}{\mbox{\tt 2+n}}}${\tt\ (2+n) 
Gamma[3+n]}$^{\frac{\mbox{\tt 2}}{\mbox{\tt 2+n}}}$

\vspace*{3mm}\hrule

\bigskip\noindent
{\tt Out[12]} is the same result as in Eq. (\ref{eq:improvede}). Let us apply Eq. 
(\ref{eq:improvede}) to typical potentials:
\begin{description}
\item[Coulomb potential:] Inserting here $a=-\alpha$ and $n=-1$, we obtain the upper bound
$$
\fbox{$E_{\rm{true}}\leq-\dfrac{\alpha^2\,\mu}{2}$}\ .
$$
Interestingly, the expression on the right-hand side is precisely the ground-state energy of a 
hydrogen atom, in other words, here the sign of equality applies.~The obvious reason for this 
clearly is that we have used the hydrogen wave function~of the ground state as our trial 
function. For $\alpha=1$ and $\mu=1$, the numerical value for $E$ is $E=-0.5$. It is very simple 
to find the minimum of Eq. (\ref{eq:totalenergy}) numerically with the help of Mathematica. 
\end{description}
\noindent\vspace*{3mm}\hrule
\begin{center}{\sl MATHEMATICA}\end{center}
\hrule\vspace*{3mm}
\noindent{\sl Defining the function to be minimized:}

\bigskip\noindent{\tt In[13]:= e[lambda\_,n\_,a\_,mu\_] := lambda\symbol{94}2/(2 mu) +
a/2 Gamma[n+3]/(2 lambda)\symbol{94}n}

\bigskip\noindent{\sl Finding the minimum (the starting point for the variational parameter 
$\lambda$ is $\lambda=0.5$):}

\bigskip\noindent{\tt In[14]:= FindMinimum[e[lambda,-1,-1,1],\{lambda,0.5\}]}

\bigskip\noindent{\tt Out[14]= \{-0.5,\{lambda}$\rightarrow${\tt 1.\}\}}

\vspace*{3mm}\hrule

\bigskip\noindent
This has to be understood as $E_{\rm{min}}=-0.5$ at the point $\lambda_{\rm{min}}=1$.

\newpage
\begin{description}
\item[Linear potential:] $V(r)=a\,r$. Inserting $n=1$ into Eq. (\ref{eq:improvede}) entails
$$
\fbox{$E_{\rm{true}}\leq E_{\rm var}=\left({\displaystyle{3\over 
2}}\right)^{5/3}\left({\displaystyle{a^2\over\mu}}\right)^{1/3}$}\ .
$$
Let us rewrite this result in the form of Eq. (\ref{eq:scaledlinear}):
\begin{equation}
E_{\rm{true}}\leq\frac{3^{5/3}}{2^{4/3}}\left(\frac{a^2}{2\,\mu}\right)^{1/3}=
2.4764\left(\frac{a^2}{2\,\mu}\right)^{1/3}\ .
\label{eq:varlin}
\end{equation}
The true ground-state energy $E_{\rm true}$ is given by the first zero of the Airy function 
\cite{abramow,lucha91},
$$
E_{\rm{true}}=2.3381\left(\frac{a^2}{2\,\mu}\right)^{1/3}\ ,
$$
which implies a surprisingly small relative error of our crude upper bound $E_{\rm var}$:
$$
\frac{E_{\rm var}-E_{\rm{true}}}{E_{\rm{true}}}\cong 6\ \%\ .
$$
We have calculated, without solving the Schr\"odinger equation, the ground-state energy for the 
linear potential to unexpectedly good approximation analytically.
\end{description}
\noindent\vspace*{3mm}\hrule
\begin{center}{\sl MATHEMATICA}\end{center}
\hrule\vspace*{3mm}
\noindent{\sl Defining the function to be minimized:}

\bigskip\noindent{\tt In[15]:= e[lambda\_,n\_,a\_,mu\_] := lambda\symbol{94}2/(2 mu) +
a/2 Gamma[n+3]/(2 lambda)\symbol{94}n}

\bigskip\noindent{\sl Finding the minimum (the starting point for the variational parameter 
$\lambda$ is $\lambda=0.5$):}

\bigskip\noindent{\tt In[16]:= FindMinimum[e[lambda,1,1,1],\{lambda,0.5\}]}

\bigskip\noindent{\tt Out[16]= \{1.96556,\{lambda}$\rightarrow${\tt 1.14471\}\}}

\vspace*{3mm}\hrule
\begin{description}
\item[{}] \mbox{}

These results may be verified by inserting $n=a=\mu=1$ into Eqs. (\ref{eq:varlin}) 
and~(\ref{eq:lambdamin}). Sometimes it is illustrative to compare plots. Letting $2\,\mu=1$, we 
shall plot, first, $E_{\rm{true}}=2.3381\,a^{2/3}$, then, $E_{\rm{var}}=2.4764\,a^{2/3}$, and, 
finally, we shall combine these two plots into one single plot with the help of the Mathematica 
command {\tt Show}.
\end{description}
\noindent\vspace*{3mm}\hrule
\begin{center}{\sl MATHEMATICA}\end{center}
\hrule\vspace*{3mm}
\noindent{\sl Defining the functions to be plotted:}

\bigskip\noindent{\tt In[17]:= etrue[a\_] := 2.3381 a\symbol{94}(2/3)}

\bigskip\noindent{\tt In[18]:= eupper[a\_] := 2.4764 a\symbol{94}(2/3)}

\newpage

\bigskip\noindent{\sl Plotting $E_{\rm true}$ ({\tt plot1}):}

\bigskip\noindent{\tt In[19]:= plot1 = Plot[etrue[a],\{a,0,3\},AxesLabel->\{"a","E"\}, TextStyle-
>\{FontSlant->"Italic",FontSize->14\}]} 

\bigskip\noindent{\sl {\tt AxesLabel} defines what shall be written on the axes; {\tt TextStyle} 
defines the font type (here italic) and the font size (here 14 pt).}

\begin{figure}[h]
\begin{center}
\psfig{figure=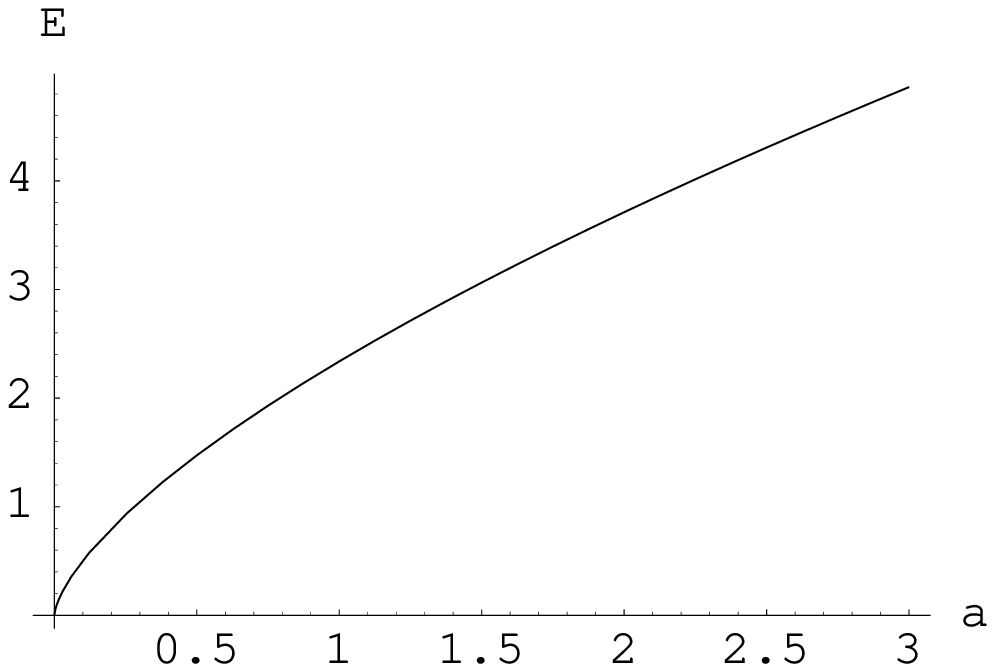, scale=1}
\caption{The true energy $E_{\rm true}$ for the ground state of the linear potential}
\end{center}
\end{figure}

\bigskip\noindent{\sl However: how is this figure---and all the others---imported into \TeX? 
After you have drawn the plot of the function using the command above, you must save the graphic 
as a {\tt ps} file. For this purpose, you write in Mathematica, after the appearence of 
the~plot:}

\bigskip\noindent{\tt Display["etrue.ps",plot1,"epsi",ImageResolution->300]}

\bigskip\noindent{\sl {\tt etrue.ps} is the name you give to the output file, {\tt plot1} is the 
name of the graphics you gave to the plot (see above) and {\tt epsi} is the necessary format for 
the graphics~output. The command {\tt ImageResolution} can be omitted, it depends on the 
resolution for the graphics you prefer. As the next step, you include your {\tt ps} file in \TeX\ 
with the help~of the following commands:}

\begin{center}
\begin{tabular}{l}
{\tt $\backslash$begin\{figure\}[h]}\\
{\tt $\backslash$begin\{center\}}\\
{\tt $\backslash$psfig\{figure=etrue.ps, scale=1\}}\\
{\tt $\backslash$caption\{etrue\}}\\
{\tt $\backslash$end\{center\}}\\
{\tt $\backslash$end\{figure\}}
\end{tabular}
\end{center}

\bigskip\noindent{\sl The command {\tt scale} enables you to resize the picture. For instance, 
{\tt scale=0.5} would make this figure precisely half as large as the original one. The 
corresponding style~file is included by the \TeX\ command\/ {\tt 
$\backslash$documentstyle[epsfig,\dots]}. In order to view~the graphics, the {\tt dvi} file has 
to be converted into a {\tt ps} file.}

\newpage

\bigskip\noindent{\sl Plotting $E_{\rm var}$ ({\tt plot2}):}

\bigskip\noindent{\tt In[20]:= plot2 = Plot[eupper[a],\{a,0,3\},AxesLabel->\{"a","E"\},
TextStyle->\{FontSlant->"Italic",FontSize->14\}]}
\begin{figure}[h]
\begin{center}
\psfig{figure=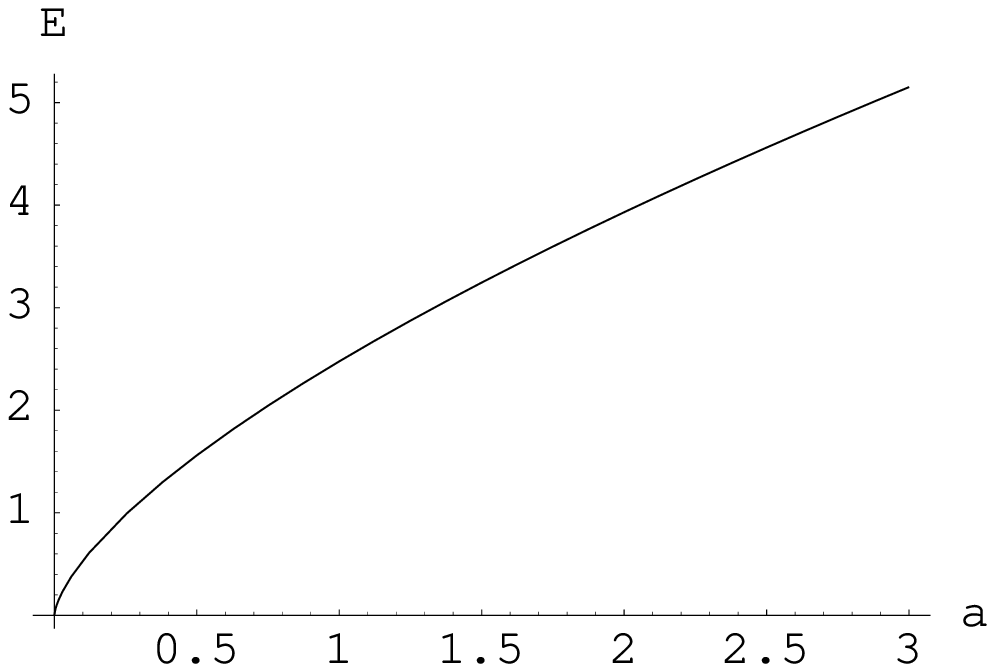, scale=1}
\caption{Variational upper bound $E_{\rm var}$ on the energy for the ground state of the linear 
potential}
\end{center}
\end{figure}

\bigskip\noindent{\sl Combining {\tt plot1} and {\tt plot2}:}

\bigskip\noindent{\tt In[21]:= Show[plot1,plot2]}
\begin{figure}[h]
\begin{center}
\psfig{figure=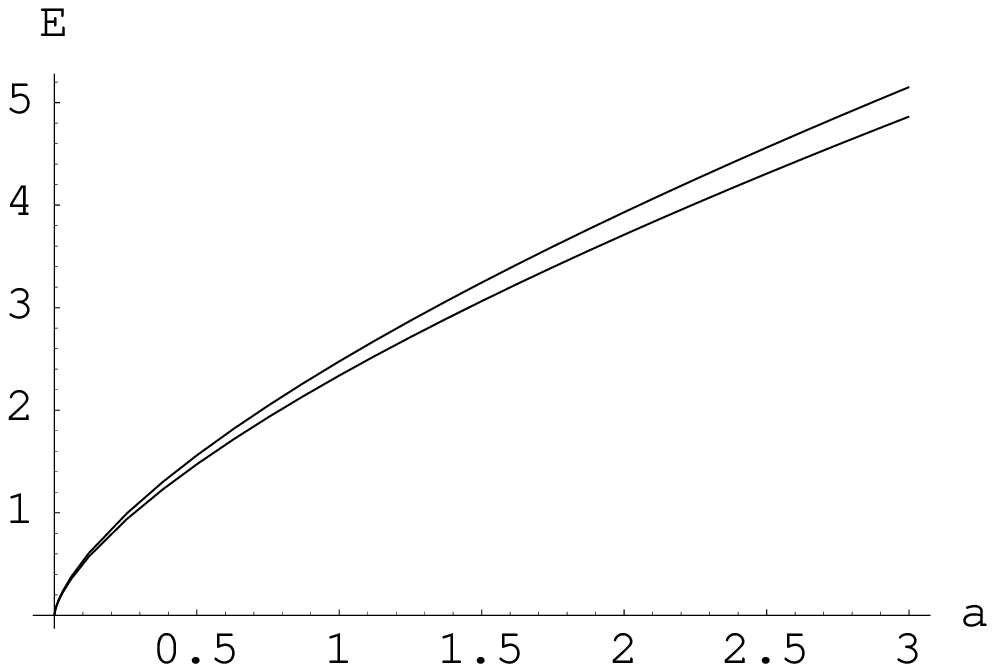, scale=1}
\caption{Exact energy and upper bound for the ground state of the linear potential}
\end{center}
\end{figure}
\vspace*{3mm}\hrule

\subsubsection{Radial Excitations}

Up to now we have discussed ground-state energies only. Next, we would~like to show how to 
determine upper bounds on energies of radially excited~states. A more detailed explanation of 
this procedure may be found in, e.g., Ref.~\cite{flamm1982}. We illustrate this method by 
determining the ground-state energy and the energy of the first radial excitation~of a system 
bound by some linear potential. First, we have to choose some trial functions. For the ground 
state, we take again the hydrogen eigenfunction
$$
\Psi_0(\lambda,r)=\frac{\lambda^{3/2}}{\sqrt{\pi}}\,e^{-\lambda\,r}\ 
,\quad\lambda^\ast=\lambda>0\ ,
$$
and, for the first radially excited state, we use
\begin{equation}
\Psi_1(\lambda,r)=\frac{\lambda^{3/2}}{\sqrt{3\pi}}\,(3-2\,\lambda\,r)\,e^{-\lambda\,r}\ 
,\quad\lambda^\ast=\lambda>0\ .
\label{eq:radialexcite}
\end{equation}
Since the radially excited states are states with nodes in the wave function, we have~to 
construct trial functions with this feature. The first radially excited state has one~node, thus 
the trial function has to have one zero, as is the case in Eq. (\ref{eq:radialexcite}). This 
trial function of the first radially excited state is constructed to be orthogonal to the 
eigenfunction~of the ground state:
$$
\left\langle\Psi_0|\Psi_1\right\rangle=0\ .
$$
To begin with, it's straightforward to check the orthogonality employing Mathematica.
\noindent\vspace*{3mm}\hrule
\begin{center}{\sl MATHEMATICA}\end{center}
\hrule\vspace*{3mm}
\noindent{\sl Defining the ground-state trial function $\Psi_0$:}

\bigskip\noindent{\tt In[22]:= psi0[lambda\_,r\_] := Sqrt[lambda\symbol{94}3/Pi] Exp[-lambda r]}

\bigskip\noindent{\sl Defining the trial function $\Psi_1$ for the first radially excited state:}

\bigskip\noindent{\tt In[23]:= psi1[lambda\_,r\_] := Sqrt[lambda\symbol{94}3/(3 Pi)] (3-2 lambda 
r) Exp[-lambda r]}

\bigskip\noindent{\sl Checking the normalization of $\Psi_0$:}

\bigskip\noindent{\tt In[24]:= 4 Pi Integrate[r\symbol{94}2 
psi0[lambda,r]\symbol{94}2,\{r,0,Infinity\}]}

\bigskip\noindent{\tt Out[24]:= 4 Pi }${\displaystyle{\tt 1\over\mbox{\tt 4 Pi}}}$

\bigskip\noindent{\sl Checking the normalization of $\Psi_1$:}

\bigskip\noindent{\tt In[25]:= 4 Pi Integrate[r\symbol{94}2 
psi1[lambda,r]\symbol{94}2,\{r,0,Infinity\}]}

\bigskip\noindent{\tt Out[25]:= 4 Pi }${\displaystyle{\tt 1\over\mbox{\tt 4 Pi}}}$

\bigskip\noindent{\sl Checking the orthogonality of the wave functions $\Psi_0$ and $\Psi_1$:}

\bigskip\noindent{\tt In[26]:= 4 Pi Integrate[r\symbol{94}2 psi0[lambda,r] 
psi1[lambda,r],\{r,0,Infinity\}]}

\bigskip\noindent{\tt Out[26]= 4 Pi 0}

\newpage

\bigskip\noindent{\sl The expectation value 
$V_{00}=\left\langle\Psi_0\right|a\,r\left|\Psi_0\right\rangle$ of the linear potential:}

\bigskip\noindent{\tt In[27]:= 4 Pi Integrate[r\symbol{94}2 psi0[lambda,r] a r 
psi0[lambda,r],\{r,0,Infinity\}]}

\bigskip\noindent{\tt Out[27]= 4 Pi} ${\displaystyle{\mbox{\tt 3 a}\over\mbox{\tt 8 lambda Pi}}}$

\bigskip\noindent{\sl The matrix elements 
$V_{01}=V_{10}=\left\langle\Psi_0\right|a\,r\left|\Psi_1\right\rangle$ of the linear potential:}

\bigskip\noindent{\tt In[28]:= 4 Pi Integrate[r\symbol{94}2 psi0[lambda,r] a r 
psi1[lambda,r],\{r,0,Infinity\}]}

\bigskip\noindent{\tt Out[28]= 4 Pi (-}${\displaystyle{\mbox{\tt Sqrt[\tt 3] a}\over\mbox{\tt 8 
lambda Pi}}}${\tt )}

\bigskip\noindent{\sl The expectation value 
$V_{11}=\left\langle\Psi_1\right|a\,r\left|\Psi_1\right\rangle$ of the linear potential:}

\bigskip\noindent{\tt In[29]:= 4 Pi Integrate[r\symbol{94}2 psi1[lambda,r] a r 
psi1[lambda,r],\{r,0,Infinity\}]}

\bigskip\noindent{\tt Out[29]= 4 Pi }${\displaystyle{\mbox{\tt 5 a}\over\mbox{\tt 8 lambda Pi}}}$

\bigskip\noindent{\sl Defining the Laplacian of the ground-state trial function $\Psi_0$:}

\bigskip\noindent{\tt In[30]:= laplacepsi0[lambda\_,r\_] := D[psi0[lambda,r],\{r,2\}]+2/r 
D[psi0[lambda,r],r]}

\bigskip\noindent{\sl Defining the Laplacian of the trial function $\Psi_1$ for the first 
radially excited state:}

\bigskip\noindent{\tt In[31]:= laplacepsi1[lambda\_,r\_] := D[psi1[lambda,r],\{r,2\}]+2/r 
D[psi1[lambda,r],r]}

\bigskip\noindent{\sl The expectation value $T_{00}=\left\langle\Psi_0\left|\dfrac{{\bf 
p}^2}{2\,\mu}\right|\Psi_0\right\rangle$ of the kinetic energy:}

\bigskip\noindent{\tt In[32]:= 4 Pi Integrate[r\symbol{94}2 psi0[lambda,r] (-
laplacepsi0[lambda,r]/(2 mu)),\{r,0,Infinity\}]}

\bigskip\noindent{\tt Out[32]= 4 Pi }${\displaystyle{\mbox{\tt lambda}^{\tt 2}\over\mbox{\tt 8 mu 
Pi}}}$

\bigskip\noindent{\sl The matrix elements $T_{01}=T_{10}=\left\langle\Psi_0\left|\dfrac{{\bf 
p}^2}{2\,\mu}\right|\Psi_1\right\rangle$ of the kinetic energy:}

\bigskip\noindent{\tt In[33]:= 4 Pi Integrate[r\symbol{94}2 psi0[lambda,r] (-
laplacepsi1[lambda,r]/(2 mu)),\{r,0,Infinity\}]}

\bigskip\noindent{\tt Out[33]= 4 Pi }${\displaystyle{\mbox{\tt lambda}^{\tt 2}\over\mbox{\tt 4 
Sqrt[3] mu Pi}}}$

\bigskip\noindent{\sl The expectation value $T_{11}=\left\langle\Psi_1\left|\dfrac{{\bf 
p}^2}{2\,\mu}\right|\Psi_1\right\rangle$ of the kinetic energy:}

\bigskip\noindent{\tt In[34]:= 4 Pi Integrate[r\symbol{94}2 psi1[lambda,r] (-
laplacepsi1[lambda,r]/(2 mu)),\{r,0,Infinity\}]}

\bigskip\noindent{\tt Out[34]= 4 Pi }${\displaystyle{\mbox{\tt 7 lambda}^{\tt 2}\over\mbox{\tt 24 
mu Pi}}}$
\vspace*{3mm}\hrule

\bigskip\noindent With the matrix elements calculated above, we are able to solve Eq. 
(\ref{eq:determinant}). However,~it is much simpler to determine the energy eigenvalues using 
Mathematica. We shall show this analytically as well as numerically.
\noindent\vspace*{3mm}\hrule
\begin{center}{\sl MATHEMATICA}\end{center}
\hrule\vspace*{3mm}
\noindent{\sl The energy matrix elements:}

\bigskip\noindent{\tt In[35]:= e00[lambda\_,mu\_,a\_] := }${\displaystyle{\mbox{\tt lambda}^{\tt 
2}\over\mbox{\tt 2 mu}}}${\tt\ + }${\displaystyle{\mbox{\tt 3 a}\over\mbox{\tt 2 lambda}}}$

\bigskip\noindent{\tt In[36]:= e11[lambda\_,mu\_,a\_] := }${\displaystyle{\mbox{\tt 7 
lambda}^{\tt 2}\over\mbox{\tt 6 mu}}}${\tt\ + }${\displaystyle{\mbox{\tt 5 a}\over\mbox{\tt 2 
lambda}}}$

\bigskip\noindent{\tt In[37]:= e10[lambda\_,mu\_,a\_] := }${\displaystyle{\mbox{\tt lambda}^{\tt 
2}\over\mbox{\tt$\sqrt{\tt 3}$}\mbox{\tt\ mu}}}${\tt\ - }${\displaystyle{\mbox{\tt$\sqrt{\tt 
3}$}\mbox{\tt\ a}\over\mbox{\tt 2 lambda}}}$

\bigskip\noindent{\sl Defining the energy matrix $E_{ij}(\lambda)$ as a function of $\lambda$, 
$\mu$, and $a$:}

\bigskip\noindent{\tt In[38]:= ematrix[lambda\_,mu\_,a\_] := 
\{\{e00[lambda,mu,a],e10[lambda,mu,a]\},
\{e10[lambda,mu,a],e11[lambda,mu,a]\}\}}

\bigskip\noindent{\sl Defining the energy eigenvalues $E(\lambda)$ as a function of $\lambda$, 
$\mu$, and $a$:}

\bigskip\noindent{\tt In[39]:= eeigen[lambda\_,mu\_,a\_] := Eigenvalues[ematrix[lambda,mu,a]]}

\bigskip\noindent{\sl Analytic evaluation of the eigenvalues:}

\bigskip\noindent{\tt In[40]:= eeigen[lambda,mu,a]}

\bigskip\noindent{\tt Out[40]= \{(5*lambda\symbol{94}4*mu + 12*a*lambda*mu\symbol{94}2 - 
2*lambda*mu*Sqrt[4*lambda\symbol{94}6 - 6*a*lambda\symbol{94}3*mu + 
9*a\symbol{94}2*mu\symbol{94}2])/(6*lambda\symbol{94}2*mu\symbol{94}2), (5*lambda\symbol{94}4*mu 
+ 12*a*lambda*mu\symbol{94}2 + 2*lambda*mu*Sqrt[4*lambda\symbol{94}6 - 6*a*lambda\symbol{94}3*mu 
+ 9*a\symbol{94}2*mu\symbol{94}2])/(6*lambda\symbol{94}2*mu\symbol{94}2)\}}

\bigskip\noindent{\sl Simplifying the above expression:}

\bigskip\noindent{\tt In[41]:= FullSimplify[\%]}

\bigskip\noindent{\tt Out[41]= \{(5*lambda\symbol{94}3 + 12*a*mu - 2*Sqrt[4*lambda\symbol{94}6 - 
6*a*lambda\symbol{94}3*mu + 9*a\symbol{94}2*mu\symbol{94}2])/(6*lambda*mu), (5*lambda\symbol{94}3 
+ 2*(6*a*mu + Sqrt[4*lambda\symbol{94}6 - 6*a*lambda\symbol{94}3*mu + 
9*a\symbol{94}2*mu\symbol{94}2]))/(6*lambda*mu)\}}

\newpage

\bigskip\noindent{\sl Numerical evaluation of the eigenvalues $E$ for $\lambda=1\ \mbox{GeV}$, 
$\mu=0.5\ \mbox{GeV}$, $a=1\ \mbox{GeV\/}^2$}:

\bigskip\noindent{\tt In[42]:= eeigen[1,1/2,1]}

\bigskip\noindent{\tt Out[42]= \{(11 - Sqrt[13])/3, (11 + Sqrt[13])/3\}}

\bigskip\noindent{\sl Numerical value (the Mathematica command\/ {\tt N({\em expr})} yields the 
numerical value of~the expression $\tt{\em expr}$):}

\bigskip\noindent{\tt In[43]:= N[\%]}

\bigskip\noindent{\tt Out[43]= \{2.46482, 4.86852\}}

\vspace*{3mm}\hrule
\bigskip\noindent The result tells us: the variational upper bounds for the energy eigenvalues 
are~given~by
\begin{eqnarray*}
E_{\rm{upper}}(\mbox{1S})&=&2.46482\mbox{ GeV}\ ,\\[1ex]
E_{\rm{upper}}(\mbox{2S})&=&4.86852\mbox{ GeV}\ ,
\end{eqnarray*}
which have to be compared with the true values (determined by the first two zeros of the Airy 
function)
\begin{eqnarray*}
E_{\rm{true}}(\mbox{1S})&=&2.33811\mbox{ GeV}\ ,\\[1ex]
E_{\rm{true}}(\mbox{2S})&=&4.08795\mbox{ GeV}\ .
\end{eqnarray*}
For the ground state, labelled 1S in usual spectroscopic notation, the relative error of our 
variational upper bound $E_{\rm{upper}}$ is
$$
\frac{E_{\rm{upper}}(\mbox{1S})-E_{\rm{true}}(\mbox{1S})}{E_{\rm{true}}(\mbox{1S})}=5.4\ \%\ ,
$$
which is, obviously, slightly better than the previous relative error of $6\ \%$. Accordingly, 
increasing the matrix size---here from $1\times 1$ to $2\times 2$---improves our upper bound. For 
the first radially excited state, labelled 2S in usual spectroscopic notation, the relative 
error~of our variational upper bound $E_{\rm{upper}}$ is
$$
\frac{E_{\rm{upper}}(\mbox{2S})-E_{\rm{true}}(\mbox{2S})}{E_{\rm{true}}(\mbox{2S})}=19.1\ \%\ ,
$$
which is, admittedly, a little bit large. However, we are able to improve these results~by 
applying Eqs. (\ref{eq:lambdamin1}) and (\ref{eq:lambdamin2}). In this way, we find the minimum 
of the energy eigenvalues covered by our chosen set of trial functions.  

However, varying the variational parameter $\lambda$ usually destroys the orthogonality~of trial 
functions attributed to different radial excitations. The characteristic equation~is, in general,  
not given by Eq. (\ref{eq:determinant}). (A detailed discussion of these troubles can be found~in 
Ref. \cite{flamm1982}.) In view of this, we focus our attention in the following to the ground 
state~1S, where this problem does not show up.

Applying the minimization procedure described at lengthy at the beginning of this section, we try 
to optimize our previously obtained upper bounds on the ground-state energy of the linear 
potential by minimizing the corresponding eigenvalue $E(\lambda)$ of our energy matrix $E_{ij}$. 
This eigenvalue is extracted from the above analytical results using the Mathematica command {\tt 
Part[{\em expr},i]} which returns the {\tt i}th part of the~expression {\em expr}.

\newpage

\noindent\vspace*{3mm}\hrule
\begin{center}{\sl MATHEMATICA}\end{center}
\hrule\vspace*{3mm}
\noindent{\sl Defining the ground-state eigenvalue as function of $\lambda$ for $\mu=0.5\ 
\mbox{GeV}$ and $a=1\ \mbox{GeV\/}^2$:}

\bigskip\noindent{\tt In[44]:= e00eigen[lambda\_] := Part[Eigenvalues[ematrix[lambda,1/2,1]],1]}

\bigskip\noindent{\sl Calculating the ground-state energy eigenvalue analytically:}

\bigskip\noindent{\tt In[45]:= e00eigen[lambda]}

\bigskip\noindent{\tt Out[45]= (6*lambda + 5*lambda\symbol{94}4 - lambda*Sqrt[9 - 
12*lambda\symbol{94}3 + 16*lambda\symbol{94}6])/(3*lambda\symbol{94}2)}

\bigskip\noindent{\sl Finding the minimum for the 1S-state energy (the starting point for 
$\lambda$ is $\lambda=0.5\ \mbox{GeV}$):}

\bigskip\noindent{\tt In[46]:= FindMinimum[\%,\{lambda,0.5\}]}

\bigskip\noindent{\sl We might expect that the correct minimum may be found employing the 
Mathematica command\/ {\tt FindMinimum[e00eigen[lambda],\{lambda,0.5\}]}. Unfortunately, for~very 
mysterious reasons, Mathematica returns a wrong result. Because of this, we carefully split the 
calculation into parts. Presumably, Mathematica is not able to simultaneously calculate  
eigenvalues, extract parts of a matrix, and find minima of some expression.}

\bigskip\noindent{\tt Out[46]= \{2.4322, \{lambda-\mbox{$>$}0.665633\}\}}
\vspace*{3mm}\hrule
\bigskip\noindent The new result, improved by minimization with respect to the variational 
parameter~$\lambda$ within the set of upper bounds obtained before, thus reads
$$
E_{\rm{var}}(\mbox{1S})=2.43220\mbox{ GeV}
$$
at the point
$$
\lambda_{\rm{min}}=0.665633\mbox{ GeV}\ .
$$
Evidently, we succeeded in reducing the corresponding relative error significantly, viz., to
$$
\frac{E_{\rm{var}}(\mbox{1S})-E_{\rm{true}}(\mbox{1S})}{E_{\rm{true}}(\mbox{1S})}=3.9\ \%\ .
$$

As mentioned before, minimization with respect to the variational parameter leads, in general, to 
different values for this variational parameter in the ground state and~in the radially excited 
states: $\lambda_i\neq\lambda_j$. Therefore, in general, all these states are no longer 
necessarily orthogonal:
$$
\left\langle\Psi_i(\lambda_i)|\Psi_j(\lambda_j)\right\rangle\neq\delta_{ij}\ .
$$
In this case, the characteristic equation for our eigenvalue problem generalizing 
Eq.~(\ref{eq:determinant}) reads
$$
\det\left[\left\langle\Psi_i(\lambda_i)\left|H\right|\Psi_j(\lambda_j)\right\rangle
-\widehat{E}\left\langle\Psi_i(\lambda_i)|\Psi_j(\lambda_j)\right\rangle\right]=0\ .
$$

In the next subsection we will show how to increase the matrix size---and therefore the accuracy 
of the computed upper bounds on the Schr\"odinger energy levels---further.

\subsubsection{Laguerre Bounds}

As mentioned before, the crucial step in obtaining accurate upper bounds is the~choice of 
convenient trial functions. Here we shall work in a basis which involves the so-called 
generalized Laguerre polynomials $L_k^{(\gamma)}$ \cite{lucha97}, specific orthogonal 
polynomials, defined by the power series \cite{abramow}
$$
L_k^{(\gamma)}(x)=\sum_{j=0}^k\,(-1)^j\left(\begin{array}{c}k+\gamma\\ k-
j\end{array}\right)\frac{x^j}{j!}
$$
and normalized according to \cite{abramow}
$$
\int\limits_0^\infty{\rm d}x\,x^\gamma\exp(-x)\,L_k^{(\gamma)}(x)\,L_{k'}^{(\gamma)}(x)= 
\frac{\Gamma(\gamma+k+1)}{k!}\,\delta_{kk'}\ .
$$
Consequently, introducing now even two variational parameters, namely, one, $\lambda$, with the 
dimension of mass as well as a dimensionless one, $\beta$, our choice for the trial function 
$\psi_{k,\ell m}({\bf x})$ with orbital angular momentum $\ell$ and projection $m$, reads in 
coordinate~space
\begin{equation}
\psi_{k,\ell m}({\bf x})={\cal N}\,|{\bf x}|^{\ell+\beta-1}\exp(-\lambda\,|{\bf x}|)\,
L_k^{(\gamma)}(2\,\lambda\,|{\bf x}|)\,{\cal Y}_{\ell m}(\Omega)\ ,
\label{eq:ansatz}
\end{equation}
where normalizability constrains the variational parameter $\lambda$ to positive values: 
$\lambda>0$. Here, ${\cal Y}_{\ell m}(\Omega)$ label the spherical harmonics for angular momentum 
$\ell$ and projection~$m$ depending on the solid angle $\Omega$; by convention, they are 
orthonormalized according to
\begin{equation}
\int{\rm d}\Omega\,{\cal Y}^\ast_{\ell m}(\Omega)\,{\cal Y}_{\ell'm'}(\Omega)= 
\delta_{\ell\ell'}\,\delta_{mm'}\ .
\label{eq:spharorth}
\end{equation}
The proper orthonormalization of our ansatz (\ref{eq:ansatz}) fixes the parameter $\gamma$ 
necessarily~to the value $\gamma=2\,\ell+2\,\beta$ and determines the normalization constant 
${\cal N}$:
$$
\psi_{k,\ell m}({\bf x})=\sqrt{\frac{(2\,\lambda)^{2\ell+2\beta+1}\,k!} 
{\Gamma(2\,\ell+2\,\beta+k+1)}}\,|{\bf x}|^{\ell+\beta-1}\exp(-\lambda\,|{\bf x}|)\,
L_k^{(2\ell+2\beta)}(2\,\lambda\,|{\bf x}|)\,{\cal Y}_{\ell m}(\Omega)
$$
satisfies the normalization condition
$$
\int{\rm d}^3x\,\psi_{k,\ell m}^\ast({\bf x)}\,\psi_{k',\ell'm'}({\bf x)}= 
\delta_{kk'}\,\delta_{\ell\ell'}\,\delta_{mm'}\ .
$$
Rather obviously, normalizability constrains the second variational parameter, $\beta$, too, 
namely, to a range characterized by $2\,\beta>-1$, i.e., to the range $\beta>-\frac{1}{2}$. For 
simplicity, we set our mass scale by choosing $m_1=m_2=1\ \mbox{GeV}$ and fix the variational 
parameters to the values $\lambda=1\ \mbox{GeV}$ and $\beta=1$. We illustrate the general 
procedure by discussing, for comparison, again the linear potential $V=r$, for simplicity with 
slope $a=1\ \mbox{GeV}^2$. At least up to a $4\times 4$ matrix, all  calculations may be 
performed analytically; we leave this to the reader as some exercise, and follow our procedure by 
applying Mathematica:
\begin{itemize}
\item define the chosen set of trial functions $\psi_{k,\ell m}({\bf x})$;
\item compute the matrix elements of the Laplacian, in other words, the kinetic energy;
\item compute the matrix elements of the radial coordinate $r$, i.e., the potential energy;
\item determine the eigenvalues $E(\lambda)$ of the resulting matrix $E_{ij}(\lambda)$ of the 
total energy;
\item compare the eigenvalues for different matrix sizes in order to observe, hopefully, 
convergence.
\end{itemize}

\newpage

\noindent\vspace*{3mm}\hrule
\begin{center}{\sl MATHEMATICA}\end{center}
\hrule\vspace*{3mm}
\noindent{\sl Defining the trial function $\psi_{k,\ell m}({\bf x})$:}

\bigskip\noindent{\tt In[47]:= psix[k\_,l\_,m\_,r\_] := Sqrt[2\symbol{94}(2 l+3) k!/Gamma[2 
l+3+k]] r\symbol{94}l Exp[-r]*LaguerreL[k,2 l+2,2 r]*SphericalHarmonicY[l,m,theta,phi]}

\bigskip\noindent{\sl Since we discuss S waves only, we define the trial function for 
$\ell=m=0$:}

\bigskip\noindent{\tt In[48]:= psi[k\_,r\_] := psix[k,0,0,r]}

\bigskip\noindent{\sl Defining the Laplacian $\Delta\psi_k(r)$ operating on states with $\ell=0$ 
(S waves):}

\bigskip\noindent{\tt In[49]:= delta[k\_,r\_] := D[psi[k,r],\{r,2\}]+2/r D[psi[k,r],\{r,1\}]}

\bigskip\noindent{\sl Defining the integrand $\psi_s(r)\,\Delta\psi_k(r)$:}

\bigskip\noindent{\tt In[50]:= intks[k\_,s\_,r\_] := psi[s,r] delta[k,r]}

\bigskip\noindent{\sl The matrix elements $\dlint_0^\infty{\rm d}r\,r^2\,\psi_s(r)\,(-
\Delta\psi_k(r))$ of the kinetic-energy operator $T=-\Delta$:\/\footnote{\ Since we have chosen 
both particle masses to be $m_1=m_2=1\ \mbox{GeV}$, twice the reduced mass~$\mu$~is also $1\ 
\mbox{GeV}$.}}

\bigskip\noindent{\tt In[51]:= kinen[k\_,s\_] := -4 Pi Integrate[r\symbol{94}2 
intks[k,s,r],\{r,0,Infinity\}]}

\bigskip\noindent{\sl The matrix elements $\dlint_0^\infty{\rm d}r\,r^2\,\psi_s(r)\,r\,\psi_k(r)$ 
of the potential-energy operator $V(r)=r$:}

\bigskip\noindent{\tt In[52]:= poten[k\_,s\_] := 4 Pi Integrate[r\symbol{94}3 psi[s,r] 
psi[k,r],\{r,0,Infinity\}]}

\bigskip\noindent{\sl The matrix elements of the total energy:}

\bigskip\noindent{\tt In[53]:= toten[k\_,s\_] := kinen[k,s]+poten[k,s]}

\bigskip\noindent{\sl Here we construct the matrix built by the above matrix elements, using the 
command {\tt Table}. Since counting of the matrix indices starts at\/ $0$, we redefine this 
matrix in~order to obtain for $x=1$ a\/ $1\times 1$ matrix, and so on.}

\bigskip\noindent{\tt In[54]:= totenmat[x\_] := Table[toten[k,s],\{k,0,x-1\},\{s,0,x-1\}]}

\bigskip\noindent{\sl Defining now the function\/ {\tt eeigen[x]} with the help of the command\/ 
{\tt Eigenvalues[{\it M}]} which gives the eigenvalues of an $x\times x$ matrix {\it M}, in other 
words, diagonalizes any $x\times x$ matrix (for our purposes, it has to be applied, of course, to 
the matrix\/ {\tt totenmat[x]}~of the total energy):}

\bigskip\noindent{\tt In[55]:= eeigen[x\_] := Eigenvalues[totenmat[x]]}

\newpage

\bigskip\noindent{\sl Eigenvalue of the\/ $1\times 1$ energy matrix:}

\bigskip\noindent{\tt In[56]:= eeigen[1]}

\bigskip\noindent{\tt Out[56]= $\left\{\dfrac{\tt 5}{\tt 2}\right\}$}

\bigskip\noindent{\sl Eigenvalues of the\/ $2\times 2$ energy matrix:}

\bigskip\noindent{\tt In[57]:= eeigen[2]}

\bigskip\noindent{\tt Out[57]= $\left\{\dfrac{\tt 1}{\tt 3}\mbox{\tt (11 - $\sqrt{\tt 13}$)},\ 
\dfrac{\tt 1}{\tt 3}\mbox{\tt (11 + $\sqrt{\tt 13}$)}\right\}$}

\bigskip\noindent{\sl Using the command\/ {\tt N[\%]} for numerical evaluation of the last 
output:}

\bigskip\noindent{\tt In[58]:= N[\%]}

\bigskip\noindent{\sl Ground-state energy and energy of the first radial excitation numerically:}

\bigskip\noindent{\tt Out[58]= \{2.46482,4.86852\}}

\bigskip\noindent{\sl Eigenvalues of the\/ $3\times 3$ energy matrix:}

\bigskip\noindent{\tt In[59]:= eeigen[3]}

\bigskip\noindent{\tt Out[59]= $\left\{\dfrac{\tt 9}{\tt 2}\mbox{\tt , 5 - $\sqrt{\tt 7}$\tt , 5 
+ $\sqrt{7}$}\right\}$}

\bigskip\noindent{\sl Using the command\/ {\tt N[\%]} for numerical evaluation of the last 
output:}

\bigskip\noindent{\tt In[60]:= N[\%]}

\bigskip\noindent{\sl Ground-state energy and energies of first and second radial excitations 
numerically:}

\bigskip\noindent{\tt Out[60]= \{4.5,2.35425,7.64575\}}

\bigskip\noindent{\sl In general, the eigenvalues of a\/ $5\times 5$ matrix cannot be found 
analytically; we determine them numerically:}

\bigskip\noindent{\tt In[61]:= N[eeigen[5]]}

\bigskip\noindent{\sl Ground-state energy and energies of the first four radial excitations:}

\bigskip\noindent{\tt Out[61]= \{2.34136,4.13334,5.72535,8.11424,15.519\}}

\bigskip\noindent{\sl The computation of the eigenvalues of the\/ $10\times 10$ energy matrix  
consumes already a noticeable amount of computer time:}

\bigskip\noindent{\tt In[62]:= N[eeigen[10]]}

\bigskip\noindent{\sl Ground-state energy and energies of the first nine radial excitations:}

\bigskip\noindent{\tt Out[62]= 
\{2.33812,4.08858,5.53209,6.83859,8.14892,9.91409,12.195,14.096,17.146, 49.7026\}}
\vspace*{3mm}\hrule

\newpage

\bigskip\noindent
Table~\ref{tab:relerror} compares the dependence of the accuracy of the obtained bounds on 
the~energy levels of ground~state and first radially excited state on the size of the energy 
matrix.\begin{table}[htb]
\caption{Comparison of relative errors of the energy eigenvalues for varying 
matrix~sizes}\label{tab:relerror}
\vspace*{3ex}
\centering
\begin{tabular}{|c|r|r|}
\hline\hline
&&\\[-1.5ex]
\multicolumn{1}{|c}{$\quad$Matrix Size$\quad$}&\multicolumn{1}{|c}{1S 
State}&\multicolumn{1}{|c|}{2S State}\\[1ex]
\hline
&&\\[-1.5ex]
$1\times 1$&6 \%$\quad$&---$\quad$\\
$2\times 2$&5 \%$\quad$&19 \%$\quad$\\
$3\times 3$&0.7 \%$\quad$&10 \%$\quad$\\
$5\times 5$&0.1 \%$\quad$&1 \%$\quad$\\
$10\times 10$&$\quad$$4\times 10^{-4}$ \%$\quad$&$\quad$$2\times 10^{-2}$ \%$\quad$\\[1ex]
\hline\hline
\end{tabular}

\end{table}

\bigskip\noindent Obviously, the accuracy obtained is very impressive. One should keep in mind 
that the above procedure is applicable not only to potentials $V(r)$ different from the power-law 
form $V=r^n$ but also to differential operators different from the nonrelativistic ${\bf 
p}^2/2\,m$. In this way one is able to determine even the energy eigenvalues of more 
sophisticated Hamiltonians like the semirelativistic one, which incorporates the square-root 
operator $\sqrt{{\bf p}^2+m^2}$ of the relativistic kinetic energy \cite{lucha97}. Moreover, it 
might be of interest to~note that up to and including $4\times 4$ matrices, the diagonalization 
of the energy matrix $E$ can be performed analytically. This fact enables one to gain control 
over purely numerically obtained results. However, one should never forget that the numerical 
values computed here represent always only upper bounds on the true energy eigenvalues: 
$E_{\rm{true}}\le E(\lambda)$.

In summary, we have tried to show how (rather complicated) bound-state problems can be handled 
``easily," using Mathematica and a few important theorems to calculate at least the resulting 
energy levels to a high precision. We didn't touch upon the task~of determining also the 
corresponding wave functions. Not surprisingly, one can show~\cite{lucha98virthm} that, for 
matrix sizes large enough, an arbitrary accuracy for the wave functions can~be achieved too. On 
the other hand, for small matrices the general conclusion is that one should not trust the wave 
functions at all, even if the energy levels are rather accurate.

For the moment, let's only produce three-dimensional plots of our trial function~for $k=0$ and 
$k=5$, to get a feeling how it looks like. The plots are scaled with {\tt scale=0.7}.

\noindent\vspace*{3mm}\hrule
\begin{center}{\sl MATHEMATICA}\end{center}
\hrule\vspace*{3mm}

\noindent{\sl Reduced trial function $\psi_{k,\ell m}({\bf x})$, to be plotted as a function 
$\psi_{k,\ell m}(x,y)$ of $x$ and $y$~only:}

\bigskip\noindent{\tt In[63]:= psix[k\_,l\_,m\_,x\_,y\_] := Sqrt[2\symbol{94}(2 l+3) k!/Gamma[2 
l+3+k]] Sqrt[x\symbol{94}2+y\symbol{94}2]\symbol{94}l Exp[-Sqrt[x\symbol{94}2+y\symbol{94}2]] 
LaguerreL[k,2 l+2,2 Sqrt[x\symbol{94}2+y\symbol{94}2]] SphericalHarmonicY[l,m,theta,phi]}

\bigskip\noindent{\sl Plotting the trial function $\psi_{k,\ell m}(x,y)$ for $k=\ell=m=0$ as a 
three-dimensional plot. This function represents by itself a suitable trial function, namely, for 
the ground~state.}

\bigskip\noindent{\tt In[64]:= Plot3D[psix[0,0,0,x,y],\{x,-4,4\},\{y,-4,4\},
TextStyle->\{FontSlant->"Italic",FontSize->12\}]}

\newpage

\begin{figure}[htb]
\begin{center}
\psfig{figure=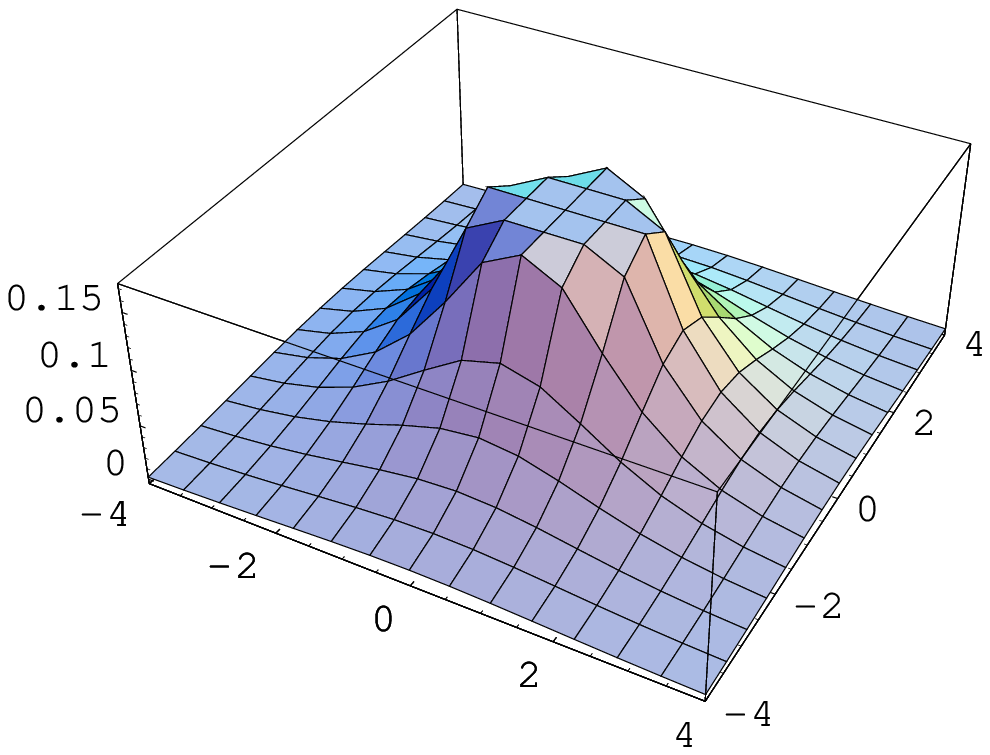,scale=0.7}
\caption{The trial function $\psi_{k,\ell m}(x,y)$ for $k=\ell=m=0$}
\end{center}
\end{figure}

\bigskip\noindent{\sl Plotting the trial function $\psi_{k,\ell m}(x,y)$ for $k=5$ and $\ell=m=0$ 
as a three-dimensional plot. For $k\neq 0$, the functions $\psi_{k,\ell m}$ are not trial 
functions corresponding to particular levels of excitation. Adequate trial functions have to be 
determined by calculating the eigenvectors of the energy matrix; they will be superpositions of 
various trial~functions.}

\bigskip\noindent{\tt In[65]:= Plot3D[psix[5,0,0,x,y],\{x,-4,4\},\{y,-4,4\},
TextStyle->\{FontSlant->"Italic",FontSize->12\}]}

\begin{figure}[htb]
\begin{center}
\psfig{figure=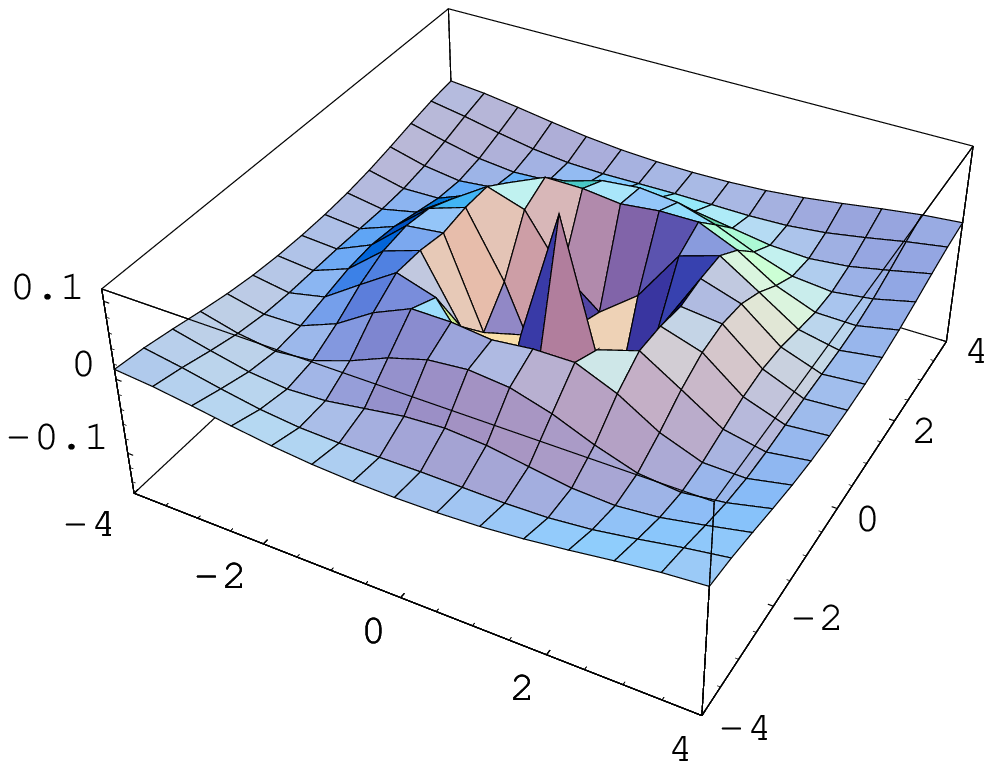,scale=0.7}
\caption{The trial function $\psi_{k,\ell m}(x,y)$ for $k=5$ and $\ell=m=0$}
\end{center}
\end{figure}
\vspace*{3mm}\hrule

\section*{Acknowledgement}

\noindent One of us, F. F. S., would like to thank the University of Science and Technology of 
China and the Institute for High Energy Physics, Beijing, of the Academy of Science of China for 
their hospitality during the stay in China.

\end{document}